\documentclass[dvips]{article}

\usepackage{icrc2011}

\title{Probing proton acceleration in W51C with MAGIC}
\newcommand{\etal}{\MakeLowercase{\textit{et al. }}} % "et al."
\shorttitle{E. Carmona \etal for the MAGIC Collaboration}

\authors{E. Carmona$^{1}$, J. Krause$^{2}$, I. Reichardt$^{3}$ for the
MAGIC Collaboration}
\afiliations{$^1$Centro de Investigaciones Energ\'eticas,
  Medioambientales y Tecnol\'ogicas, Madrid, Spain\\ $^2$Max-Planck-Institut f\"ur Physik, Munich, Germany  \\$^3$Institut de F\'isica d'Altes Energies, Barcelona, Spain }
\email{emilianocarm@googlemail.com}

\abstract{
Located in a dense complex environment, W51C provides an excellent scenario to probe
accelerated protons in SNRs and their interaction with surrounding target material.
Here we report the observation of extended Very High Energy (VHE) gamma-ray emission
from the W51C supernova remnant (SNR) with MAGIC. Detections of extended gamma-ray
emission in the same region have already been reported by the Fermi and H.E.S.S.
collaborations. {\em Fermi}/LAT measured the source spectrum in the energy range between 0.2
and 50 GeV, which was found to be well fit by a hadronic neutral-pion decay model.
The VHE observations presented here, obtained with the improved MAGIC stereo system,
allow us to pinpoint the VHE gamma-ray emission in the dense shocked molecular cloud
surrounding the remnant shell. The MAGIC data also allow us to measure, for the first
time, the VHE emission spectrum of W51C from the highest {\em Fermi}/LAT energies up to several
TeV. The spatial distribution and spectral properties of the VHE emission suggest a
hadronic origin of the observed gamma rays. Therefore W51C is a prime candidate for a
cosmic ray accelerator.
}

\begin{document}
\maketitle

%Begin the section.
\section{Introduction}
%Since cosmic rays were first discovered, the question about their
%origin has remained unanswered. 
For many years Supernova Remnants
(SNRs) have been suggested as one of the main locations for the
acceleration of cosmic rays, at least for those originated within the
Galaxy. From a theoretical point of view, cosmic rays can be
accelerated in the expanding shocks of the SNRs, receiving part of the
kinetic energy of the shock through the diffusive shock acceleration
mechanism~\cite{Reynolds2008}. The last years have provided increasing
observational evidence supporting this idea. The observations of high
energy (HE) and very high energy (VHE) gamma rays from many SNRs might
be the proof that cosmic rays are accelerated in SNRs, if the hadronic
origin of these gamma rays is unambiguously established.

In most cases, electromagnetic scenarios (gamma rays produced by
up-scattering of photons through Inverse Compton scattering by
accelerated electrons) and
hadronic scenarios (gamma rays produced after the decay of $\pi^0$)
can not be distinguished by looking only at the VHE gamma-ray
emission. Observations at other wavelengths are needed to better
understand the production mechanism of gamma rays. In particular, the
observations of HE gamma rays in the range from MeVs to GeVs are very
important because they can reveal features in the spectrum
that can distinguish both possibilities.

Evidence of cosmic rays acceleration is provided by gamma rays emitted
by molecular clouds adjacent to or shocked by
SNRs~\cite{Hinton2010}. VHE gamma-ray emission from
W28~\cite{HESS-W28} and IC443~\cite{MAGIC-IC443} are two examples
where this effect might be responsible for the production of VHE
gamma rays. In the W28 case, gamma rays might be produced by cosmic
rays that escaped the accelerating SNR while in IC 443
a {\em Crushed Cloud} scenario might be at work where the molecular cloud is shocked by the
blastwave region of the SNR~\cite{Uchiyama2010}. 

The SNR W51C offers an excellent opportunity to study the emission of
VHE gamma rays in a {\em Crushed Cloud} scenario. W51C is a composite
SNR with an elliptical shape and a size of $0.8^\circ \times
0.6^\circ$.  It is located in the tangential point of the Sagittarius
arm at a distance of $\sim6$~kpc~\cite{Koo1995} and the estimated age
is around 30~kyrs. 
The W51 complex hosts three main components: Two star-forming
regions, W51A and W51B, and the SNR W51C. While W51A is separated from
the other two, W51B overlaps with the North-Western rim of W51C. 
Shocked atomic and molecular gases have
been observed in radio data~\cite{Koo1997a}~\cite{Koo1997b}, providing
direct evidence on the interaction of the W51C shock with a large
molecular cloud. X-ray emission has been detected from the star
forming regions in W51A and W51B. W51C is also visible in X-rays
showing both a shell type and center-filled morphology~\cite{Koo2002}. Non-thermal
X-ray emission has also been detected from the relatively bright
source CXO J192318.5+140505, which is thought to be a pulsar wind
nebula (PWN) associated to the SNR~\cite{Koo2005}.

{\em Fermi}/LAT detected gamma-ray emission from 200MeV to 50GeV
extended throughout W51B and W51C~\cite{FermiW51}. The relatively
large PSF does not allow to tell from which of the objects of the
field of view the emission comes from. The {\em Fermi}/LAT emission
region is extended in comparison to the PSF of the
instrument. H.E.S.S. has also detected VHE emission above 420~GeV from
an extended region coincident with the {\em Fermi}/LAT emission
region~\cite{HESS-W51}. In their skymap, the H.E.S.S. emission is
smoothed with a radius of 0.22$^\circ$~\cite{HESS-W51} and, as well as
in the case of {\em Fermi}/LAT, it overlaps with several HII regions,
the molecular mass in W51B as well as the PWN candidate
CXOJ192318.5+140305. The flux measured by H.E.S.S. above 1~TeV is at
the level of 3\% of the Crab Nebula flux. Finally, also the Milagro
Collaboration reported a possible excess from the same source at
energies above several TeV~\cite{Milagro-W51}.

The modelling of the spectral energy distribution (SED) measured by
{\em Fermi}/LAT strongly points to a hadronic mechanism as the main origin
of the gamma rays. The interaction should take place in the region
between the supernova remnant and the dense, complex region North-West
of it. This emission might be produced by the {\em Crushed Cloud}
scenario~\cite{Uchiyama2010}, where preexisting cosmic rays are
accelerated after the passage of the supernova blast wave. In this
case, the cloud engulfed would be those described in~\cite{Koo1997a}\cite{Koo1997b}.

\section{Observations with the MAGIC telescopes}

MAGIC consists of two 17~m diameter imaging atmospheric Cherenkov
telescopes  located at the Roque de los Muchachos in the Canary Island
of La Palma ($28^{\circ}46'$N, $17^{\circ}53'$W) at the height of
2200~m\,a.s.l. Astronomical observations of very high energy gamma-ray sources are
performed with the two telescopes simultaneously which provides a major
improvement of performance with respect to the single telescope
observations previously done with MAGIC-I~\cite{StereoPaper}.

MAGIC observed W51C between May 17 and August 19 2010. The
observations were carried out in the so-called wobble mode and covered
a zenith angle range between 14 and 35 degrees. As central position
for the observations, the center of the {\em Fermi}/LAT source W51C (RA=19.385~h,
$\delta$=14.19~$^\circ$) was chosen. After applying quality cuts we
collected a total of 31.1~h effective\footnote{Due to deadtime of the
  MAGIC-II telescope readout system the effective observation time is
  always lower than the real observation time.} dark time.  All data
was taken in stereoscopic mode were only events that triggered both
telescopes are stored.

The trigger energy threshold of the system is around
50~GeV~\cite{StereoPaper}. This is the lowest energy threshold among
IACTs in operation and provides the chance to have ground based
observations with an energy range overlapping with that of {\em Fermi}/LAT.
The analysis of the data was performed using the MARS analysis framework which
is the standard software used for MAGIC data
analysis~\cite{MoralejoLodz}. The details of the analysis, as well as
the general performance of MAGIC in stereoscopic mode, are reported
in~\cite{StereoPaper}~\cite{StereoICRC}.

\section{Results from MAGIC observations}

\begin{figure}[!t]
  \vspace{5mm} \centering
  \includegraphics[width=0.95\columnwidth]{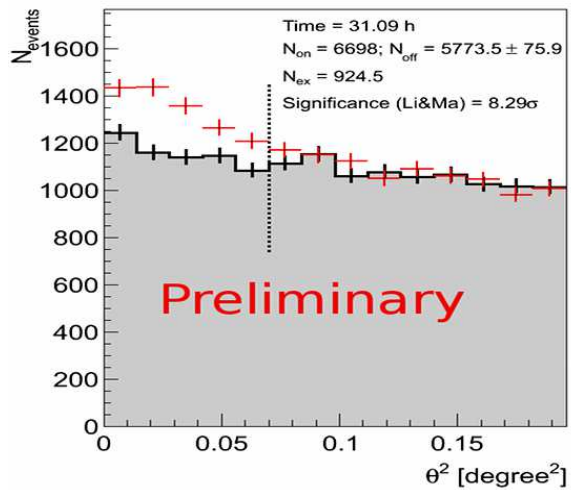}
  \caption{Distribution of the squared angular distance between the
    reconstructed VHE gamma-ray events (red
    points) and background (black points) above 150~GeV after cuts, and
    the source position. The vertical dotted line marks the
  region defined to compute the excess events. }
  \label{Theta2}
 \end{figure}

\begin{figure}[!t]
  \vspace{5mm}
  \centering
  \includegraphics[width=0.95\columnwidth]{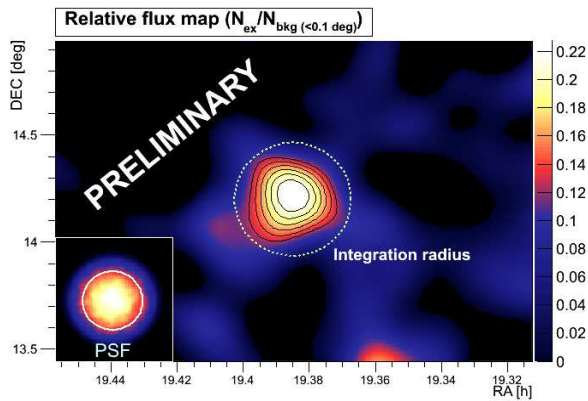}
  \caption{VHE gamma-ray emission from W51C obtained with the MAGIC
    telescopes above 150 GeV. The map has been smoothed with a
    gaussian kernel of $\sigma$=0.10$^\circ$. The color scale shows
    the relative flux of gamma rays (excess events normalized to the
    number of background events) and the black contours are different
    levels of the test statistic variable (Li \& Ma
    eq. 17~\cite{LiMa83} applied on a smoothed and modelled background
    estimation. It roughly corresponds to a gaussian significance.
    More details given in~\cite{Saverio2011}). In the figure the region
    defined for integrating the signal (dotted line) and the PSF after
    smearing (bottom left inset) are also shown.}
  \label{SkymapMagic}
 \end{figure}

MAGIC data show a spatially extended excess of VHE gamma-ray events
from the direction of W51C. A total of 924 excess events were found above
150~GeV in the analyzed  31.1~hours of 
effective time (see figure~\ref{Theta2}). This excess
corresponds to a significance of 8.29~$\sigma$. The source has an
extension (sigma of two-dimensional gaussian fit) of
0.16$\pm$0.02$^\circ$, well above that of the point spread function (PSF) of
the analysis (0.08$^\circ$ in the same energy range from a spectral
index of -2.6). The centroid of the emission detected by MAGIC is
coincident with the centroid of the {\em Fermi}/LAT emission:
RA=19.387$\pm$0.002~h, $\delta$=14.18$\pm$0.02$^\circ$. A skymap of
the MAGIC observation at energies above 150~GeV is shown in
figure~\ref{SkymapMagic}. Details about MAGIC skymaps and test
statistic are given in~\cite{Saverio2011}.

\begin{figure}[!t]
  \vspace{5mm}
  \centering
  \includegraphics[width=0.95\columnwidth]{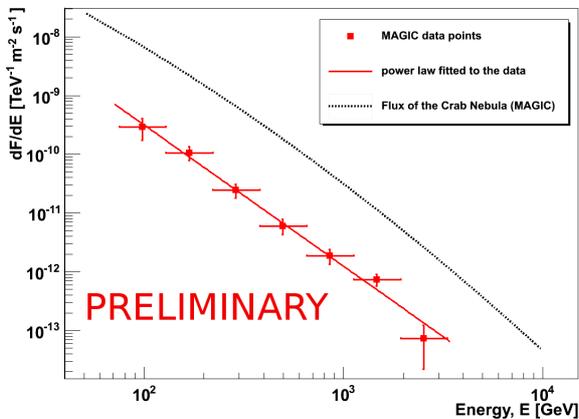}
  \caption{Measured flux from W51C with the MAGIC telescopes. The red
    points show the differential fluxes. The red line represents the
    best fitted power-law. As a reference the measured by MAGIC-I
    differential energy spectrum from the Crab Nebula is also shown.}
  \label{Spectrum}
 \end{figure}

MAGIC has measured the differential energy spectrum of the VHE
gamma-ray emission in the energy range of 75~GeV to 3.3~TeV. The
measured differential spectrum is well fitted by a power law as can be
seen in figure~\ref{Spectrum} ($\chi^2/ndf$ = 4.5/5). The
obtained spectral index of the  VHE gamma-ray emission is
-2.40$\pm$0.12$_{stat}$. The flux at 1 TeV corresponds to 3.8\% of the
Crab Nebula in agreement with the integral flux reported by
H.E.S.S.~\cite{HESS-W51} above
1 TeV. The obtained flux in units of TeV$^{-1}$ cm$^{-2}$ s$^{-1}$
is given by: 

\begin{eqnarray*}
\frac{dF}{dE} = \mathrm {(1.25 \pm 0.18_{ stat}) \times 10^{-12}}
\left ( \frac{E}{TeV}\right ) ^ \mathrm{(-2.40\pm0.12_{stat})}
\end{eqnarray*}

\section{Discussion}

\begin{figure}[!t]
  \vspace{5mm}
  \centering
  \includegraphics[width=0.95\columnwidth]{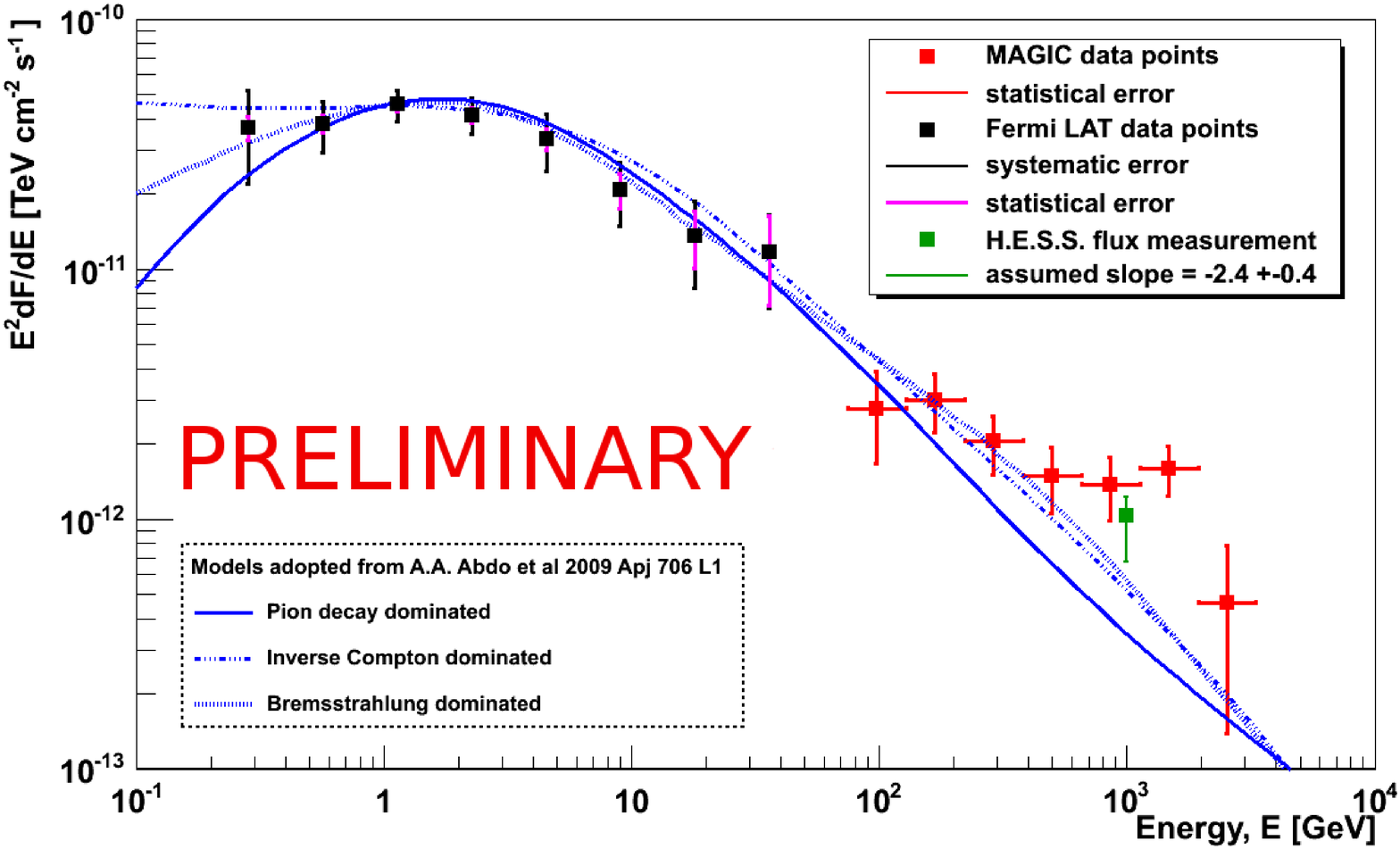}
  \caption{SED of W51C measured by {\em Fermi}/LAT (black points), MAGIC (red
    points) and H.E.S.S. (green point). Also shown in the figure the 3
  different scenarios for modelling the multiwavelength data that were
  used in~\cite{FermiW51}. Gamma-ray emission is explained by
  emission coming from a hadronic dominated scenario (continuous blue
  line), inverse Compton dominated scenario (dot-dashed line) or a
  Bremhstrahlung dominated scenario (dashed line). }
  \label{SED}
 \end{figure}

The MAGIC data fill the gap between the {\em Fermi}/LAT and the
H.E.S.S. measurements. Figure~\ref{SED} shows the SED measured by
{\em Fermi}/LAT and MAGIC together with the H.E.S.S. measurement converted into
a differential flux\footnote{The integral flux above 1 TeV reported
  in~\cite{HESS-W51} is converted into a differential flux using
  the MAGIC spectral index of -2.4. An error of $\pm0.4$ is assumed in
  order to obtain the error bars shown.}. MAGIC data
agrees well with the {\em Fermi}/LAT and H.E.S.S. measurements. In the same
figure the predictions from the phenomenological model used by {\em Fermi}/LAT
in~\cite{FermiW51} to explain the origin of the gamma-ray emission are
also shown. Three different scenarios are considered: one where
gamma-ray emission is dominated by $\pi^0$ decay and two more where
the gamma-ray emission is mainly dominated by inverse Compton or
Brehmstrahlung.

VHE MAGIC data points as well as H.E.S.S. point seem to be slightly
above predictions from the model in all scenarios at energies above 1
TeV. Although this seems to favour the electromagnetic scenarios, it
may also be due to the spectra of pre-existing cosmic rays in the
cloud being different from the one assumed in the model (the galactic
cosmic ray spectrum). In addition, electromagnetic scenarios have
problems to fit the radio data and they need a ratio of radiating
electrons to protons in the SNR shock very far from what it is
observed in cosmic rays. Taken into account all this, the hadronic
scenario is the most likely one to explain the origin of the HE-VHE
emission.

\begin{figure*}[th]
  \centering
  \includegraphics[width=6in,height=3in]{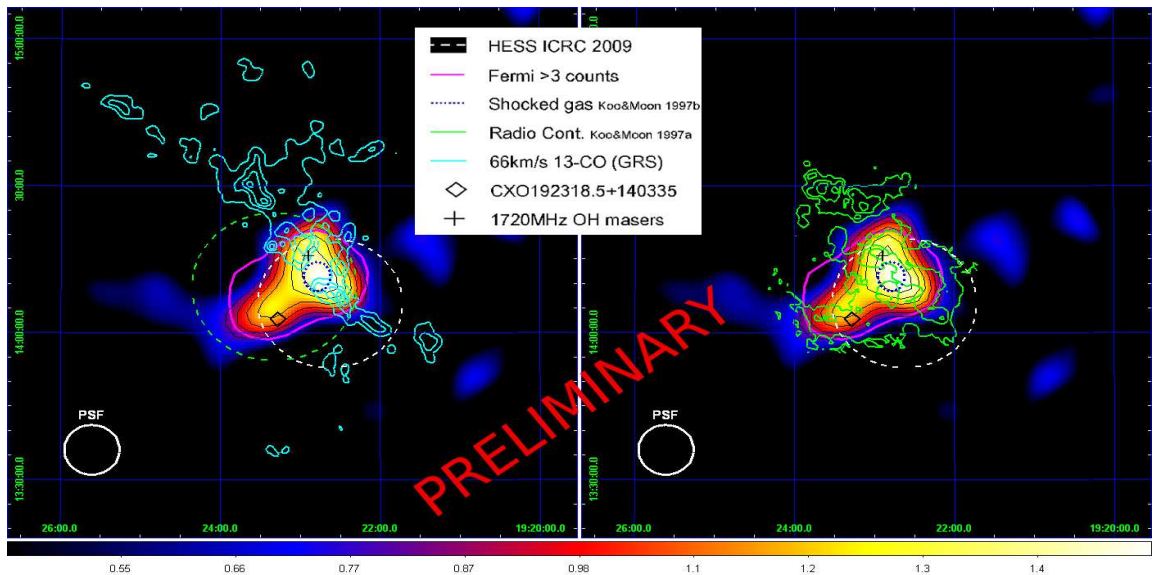}
  \caption{Map of W51C region in different wavelenghts. Shown in
    colors the MAGIC relative flux (smooth by a gaussian kernel of
    0.065$^\circ$) overlapped with the test statistics significance
    contours in black. The pink line shows the approximate contour of
    the HE emission detected by {\em Fermi}/LAT and the white dashed line shows
    the approximate contour of the H.E.S.S. VHE emission. The green
    dashed line shows the approximate contour of the SNR W51C. The
    dotted dark blue line shows the shocked gas region defined by Koo
    et al.~\cite{Koo1997b}. On the left map the molecular clouds measured with the
    66~km/s 13-CO line are shown in light blue. On the right map, the
    green contours show the radio data from Koo \&
    Moon~\cite{Koo1997a} in green. The X-ray source and PWN candidate
    CXO192318.5+140335 and a 1720MHz OH maser are also shown in the
    map.}
  \label{MW_Map}
 \end{figure*}

The angular resolution of MAGIC at energies of 100~GeV is comparable
to that of {\em Fermi}/LAT. At energies above 700~GeV, however, the
angular resolution of the MAGIC stereo system is
$\simeq0.05^\circ$. This allows for a higher-resolved skymap from
W51C. Figure~\ref{MW_Map} shows the MAGIC view of W51C overlapped with
data from different wavelengths. The bulk of the VHE gamma-ray
emission above 700~GeV is coincident with the shocked gas region
reported by Koo et al. in~\cite{Koo1997b}. There is also an extension
of the VHE emission towards the South-East in the direction of the PWN
candidate, following the shape of the {\em Fermi}/LAT HE
emission. This may be an indication for an additional, harder emission
component from the PWN, that might explain also the differences with
respect to the models at VHE in the SED. MAGIC emission above 700~GeV
is contained within the region defined by {\em Fermi}/LAT (pink line
in the figure), but clearly favours the origin of the gamma-ray
emission to be located in the shocked gas region rather than the PWN
candidate. This also supports the hadronic origin of the HE-VHE gamma
rays.

\section{Conclusions}

MAGIC observations confirm the emission of HE-VHE gamma rays from a
extended source located in the SNR W51C. The emission measured by
MAGIC is spatially coincident with that reported by {\em Fermi}/LAT, and the
measured SED is in agreement with the {\em Fermi}/LAT and
H.E.S.S. measurements. Moreover, the higher angular resolution
provided by MAGIC shows that the bulk of the VHE gamma-ray emission
comes from the shocked molecular cloud located where the SNR shock
engulfs a large molecular cloud, creating a shocked gas region
distinguishable in the radio data. This fact, and the better agreement
of radio data with the hadronic scenario, suggests that the gamma-ray
emission has most likely  a hadronic origin as it is expected in the
case of gamma rays produced in a {\em crushed cloud} scenario.

%\section*{Acknowledgements}
%We would like to thank the Instituto de Astrof\'{\i}sica de
%Canarias for the excellent working conditions at the
%Observatorio del Roque de los Muchachos in La Palma.
%The support of the German BMBF and MPG, the Italian INFN, 
%the Swiss National Fund SNF, and the Spanish MICINN is 
%gratefully acknowledged. This work was also supported by 
%the Marie Curie program, by the CPAN CSD2007-00042 and MultiDark
%CSD2009-00064 projects of the Spanish Consolider-Ingenio 2010
%programme, by grant DO02-353 of the Bulgarian NSF, by grant 127740 of 
%the Academy of Finland, by the YIP of the Helmholtz Gemeinschaft, 
%by the DFG Cluster of Excellence ``Origin and Structure of the 
%Universe'', and by the Polish MNiSzW Grant N N203 390834.

%%%\vspace{\baselineskip}

\clearpage

\end{document}